\documentclass{aa}  
\usepackage{graphicx}
\usepackage{txfonts}
\begin{document}
\title{Discovery of a bright eclipsing cataclysmic variable}

\author{D. K. Sing\inst{1,2}, E. M. Green\inst{3}, S. B. Howell\inst{4}, J. B. Holberg\inst{2}, M. Lopez-Morales\inst{5}\thanks{Carnegie Fellow}, J. S. Shaw\inst{6}, G. D. Schmidt\inst{3}}

   \offprints{D. K. Sing}

\institute{Institut d'Astrophysique de Paris, CNRS/UPMC, 98bis boulevard Arago, 75014 Paris, France\\ \email{sing@iap.fr}
 \and
Lunar and Planetary Laboratory, Sonett Bld., University of Arizona, Tucson, AZ 85721, USA\\ \email{holberg@argus.lpl.arizona.edu}
 \and
Steward Observatory, University of Arizona, 933 North Cherry Avenue, Tucson, AZ 85721, USA\\ \email{bgreen@as.arizona.edu,gschmidt@as.arizona.edu}
 \and
WIYN Observatory \& NOAO, P.O. Box 26732, 950 N. Cherry Ave., Tucson, AZ 85726-6732, USA\\ \email{howell@noao.edu}
 \and
Carnegie Institution of Washington, Dept. of Terrestrial Magnetism, 5241 Broad Branch Road NW, Washington, DC 20015, USA\\ \email{mercedes@dtm.ciw.edu}
 \and
Department of Physics and Astronomy, University of Georgia, Athens GA, 30602, USA \\ \email{jss@hal.physast.uga.edu}}

   \date{Received; accepted }

 
  \abstract
   {}
   {We report on the discovery of J0644+3344, a bright deeply eclipsing cataclysmic variable (CV) binary.}
   {Optical photometric and spectroscopic observations were obtained to determine the nature and characteristics of this CV.}
   {
Spectral signatures of both binary components and an accretion disk can be
seen at optical wavelengths.   The optical spectrum shows broad H I,
He I, and He II accretion disk emission lines with deep narrow
absorption components from H I, He I, Mg II and Ca II.
The absorption lines are seen throughout the orbital period, disappearing only
during primary eclipse.  These absorption lines are either the the
result of an optically-thick inner accretion disk or from
the photosphere of the primary star.  Radial velocity measurements
show that the H I, He I, and Mg II absorption lines phase with the
the primary star, while weak absorption features in the
continuum, between H$\alpha$ and H$\beta$, phase with the secondary
star.  Radial velocity  solutions give a 150$\pm$4 km s$^{-1}$
semi-amplitude for the primary star and 192.8$\pm$5.6 km s$^{-1}$ for
the  secondary, resulting in a primary to secondary mass ratio of
$q$=1.285.  The individual stellar masses are 0.63$-$0.69  
M$_{\odot}$ for the primary and 0.49$-$0.54 M$_{\odot}$ for the
secondary, with the uncertainty largely due to the inclination.  
}
   {
The bright eclipsing nature of this binary has helped provide masses for
both components with an accuracy rarely achieved for cataclysmic variables.
This binary most closely resembles a nova-like UX UMa or SW Sex type of
cataclysmic variable.  J0644+3344, however, has a longer orbital period
than most UX UMa or SW Sex stars.   Assuming an evolution toward shorter orbital
periods, J0644+3344 is therefore likely to be a young interacting binary.
The secondary star is consistent with the size and spectral type of a K8 star, 
but has the mass of a M0. 
}

\titlerunning{Discovery of a bright eclipsing cataclysmic variable}
\authorrunning{Sing et al.}
   \maketitle
%

\section{Introduction}

Cataclysmic Variables (CVs) are close binary systems consisting of a
white dwarf (WD) primary and late type secondary which overflows its Roche
lobe.  In these systems, mass is transfered from the secondary onto
the primary star, often forming an accretion disk.   CVs are
thought to have been produced through the common-envelope (CE)
process.  In the CE process, one component of a binary
star system has a rapidly expanding  envelope, evolving on either the
red giant branch (RGB) or the asymptotic giant branch (AGB).  The
giant fills its Roche lobe starting rapid mass transfer, quickly
filling its companion's Roche lobe as well.   The companion star
spirals in helping to eject the envelope which has formed around both
stars.  The initial conditions of the binary and the efficiency of
ejecting the envelope determine the final separation of the binary.
After the ejection of the envelope, the core of the giant is left
behind which has formed into either a subdwarf or white dwarf, while
the secondary's mass remains nearly unchanged.  Those post-CE systems
which have not coalesced are thought to then go through a detached
phase, where further orbital angular momentum is lost through
gravitational radiation and magnetic braking.  Post-common envelope
systems that have evolved to the point where the secondary's
Roche-lobe is filled, either by the secondary evolving off of the main
sequence, or by loss of orbital angular momentum and favorable
mass ratios, then become CVs.   There are approximately 28 known
post-common envelope systems which are also believed to be pre-CVs
(Sing 2005).

Although the accretion disk usually dominates the optical continuum out-shining the primary, the white
dwarf primary can be observed if it is sufficiently hot or in cases of
low mass transfer.  The masses, radii, and spectral line profiles
measured for the primary star in CVs have thus far all been consistent
with white dwarf characteristics.  Certain nova-like (NL) CVs were
previously thought to contain a hot subdwarf, such as UX UMa (Walker
\& Herbig 1954), because of the appearance of broad absorption lines.
These CVs, however, were shown to contain optically thick inner
accretion disks which gives rise to the broad absorption features.

We present discovery observations of a new bright NL CV 
NSVS07178256 (hereafter J0644).  
We identified this system as a variable star while conducting an extensive search campaign for short
period eclipsing binaries among objects in the Northern Sky
Variability Survey (NSVS, Wozniak 2004) photometry database.  The
coordinates for J0644 are included in Table 1 along with a finder
chart in Fig.\ 1.
The NSVS database was searched for periodicities by simultaneously
using  the String-Length (Clarke 2002) and Analysis of Variance
(Schwarzenberg-Czerny 1989) methods. J0644 revealed itself as a
possible candidate having strong photometric variations exceeding one
magnitude, interpreted  as brief periodic eclipses occurring with a
period of 0.538669 days.

\begin{figure}
  \includegraphics[width=9cm]{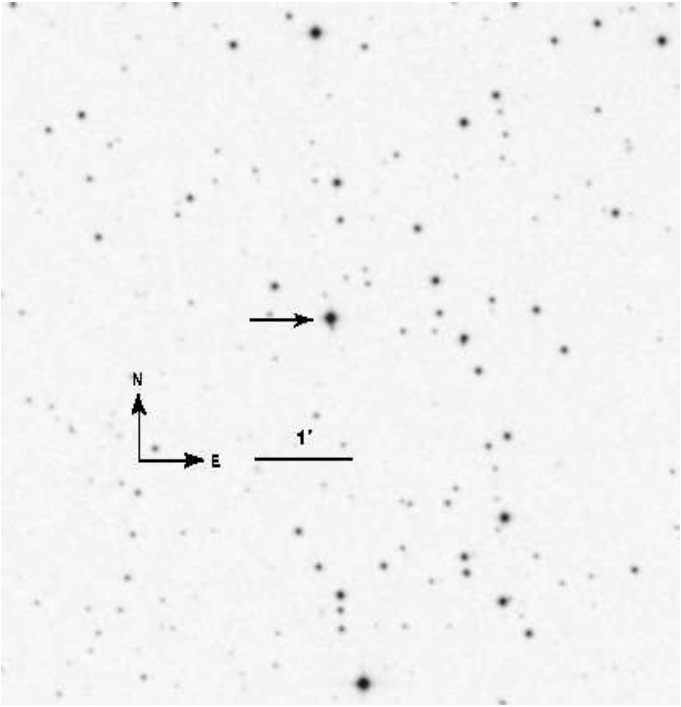}
  \caption[J0644.finder.ps]{Digital Sky Survey (Poss II, blue) finding chart for J0644+3344.}
\end{figure}

There are no previous references to this binary in the literature
outside of our initial observations (Sing 2005).  The only initial
estimation that we had of its apparent magnitude was the  I-band value
provided by the NSVS ($I_{NSVS} \sim 13.4$). A more  careful search on
this object revealed that it can be identified with  the ROSAT All Sky
Survey source 1RXS J064434.5+334451. The object also  appears in the
2MASS catalog as object 2MASS J06443435+3344566 (J=12.493 $\pm$ 0.023;
H=12.163 $\pm$ 0.026; Ks=12.03 $\pm$ 0.022). The  J-Ks, J-H, and H-Ks
colors from 2MASS (J-Ks=0.463 $\pm$ 0.032; J-H=0.330 $\pm$ 0.035, and
H-Ks = 0.133 $\pm$ 0.034) provide the first clues about the nature of
this binary.  From the UCAC2 catalog (Zacharias et. al.\ 2004), we
find that J0644 has a  small proper motion ($\mu_a$= -5 $\pm$
4.3; $\mu_d$= -3.3 $\pm$ 1.7)

Its H-Ks color is consistent with the system being a low-mass
eclipsing binary, however its J-H color is too blue due to the 
disk contribution (J-H $>$ 0.5 for
low-mass stars). Placing this binary in the (J-H) vs  (H-Ks) diagrams
for CVs (Hoard et al.\ 2002) suggested that J0644 had the
characteristics of a NL CV.

We obtained new photometric and spectroscopic observations of
J0644 in an attempt to identify the nature of the binary system,
refine its orbital period, and perform radial velocity
measurements to allow estimation of the component masses.  We describe these observations in \S 2, analyze the
data in \S 3, discuss the results in \S 4, and present our
conclusions in \S 5.

\begin{table} 
\caption{J0644+3344 System Parameters}
\label{Table 1}
\centering
\begin{tabular}{lll}
\hline\hline
Parameter & Value & Value \\
\hline
RA(2000)                   &  06:44:34.637                    &                       \\
DEC(2000)                  &  +33:44:56.615                   &                       \\
U                          &  13.11 $\pm$0.07                  &                       \\
B                          &  13.84 $\pm$0.05                  &                       \\
V                          &  13.56 $\pm$0.24                  &                       \\
R                          &  13.29 $\pm$0.05                 &                       \\
I                          &  12.8 $\pm$0.1                  &                       \\
J                          &  12.493$\pm$0.023                & 2MASS                      \\
H                          &  12.163$\pm$0.026                & 2MASS                      \\
K                          &  12.03 $\pm$0.022                & 2MASS                      \\
ROTSE magnitude            &  13.392                          & Unfiltered            \\
Secondary                  & K8 (Spectral Type)               & M0 (Mass)         \\
$K_{P}$ Velocity          & 150 $\pm$ 4 km s$^{-1}$           & He II solution    \\
$K_{2}$ Velocity          & 192.8 $\pm$ 5.6 km s$^{-1}$      &                   \\
$\gamma$ system            &  -7.1 $\pm$ 1.3 km s$^{-1}$      &                   \\                  
$M_{P}(i)$                 &  0.63$-$0.69 M$_{\odot}$           & Mass range       \\
$M_{2}(i)$                &  0.49$-$0.54 M$_{\odot}$           & Mass range       \\  
$a(i)$                     &  1.82-1.88 R$_{\odot}$           &                   \\
$i$                        & $>$ 76$^{\circ}$                 & Inclination        \\
\hline
\end{tabular}
\end{table}

\section{Observations}

\subsection{Kuiper 1.55m Photometry} 

Time resolved differential photometry of J0644+3344 (Table~2) was
obtained in February, March, November and December of 2005 and in
January and October of 2006 at the Steward Observatory 1.55m (61" Kuiper)
Telescope located on Mt.\ Bigelow.  Most of the early 2005
observations were taken with a NSF Lick3 2048$\times$2048 pixel CCD,
for which the image scale of 0.15$^{\prime\prime}$ pixel${-1}$ allowed
a 5.1$^{\prime}$$\times$5.1$^{\prime}$ field of view.  A new facility
imager with a blue-sensitive dual-amplifier Fairchild 4096$\times$4096
CCD was introduced in September 2005, expanding the available field of
view to 10.2$^{\prime}$$\times$10.2$^{\prime}$.  With both
instruments, 3$\times$3 pixel on-chip binning was utilized to
substantially improve the readout time.  The average overhead times
between successive images in a single filter were 26~s and 23~s for
the 2K and 4K CCD's, respectively.  Each filter change required an
additional 10~s with the 2KCCD, or 6~s with the 4KCCD.  On the nights
for which two filters are listed in Table~2, the filters were
alternated for the entire observation period.  The sampling times can
be calculated using the overhead values plus the exposure times in
Table~2.  For example, the sampling time was 71~s for the 04~Feb 2005
data, and it was 133~s between two exposures in the same filter on
06~Jan 2006.

The data were reduced using standard IRAF\footnote{The Image Reduction
and Analysis Facility, a general purpose software package for
astronomical data, is written and supported by the IRAF programming
group of the National Optical Astronomy Observatories (NOAO) in
Tucson, AZ.} routines for bias subtraction, flat-fielding, and cosmic
ray cleaning.  Precise differential photometry was obtained using
IRAF's APPHOT package to measure the magnitude of J0644+3344 relative
to the mean magnitudes of a set of eight to ten comparison stars in
each image.  The size of the stellar aperture radius (2.25 times the
observed stellar FWHM) and the background sky annulus (four times the
area of the star aperture with an inner boundary of 5.0 times the
stellar FWHM) were held constant for all stars in a given image,
although the values varied from image to image, according to the
seeing.  An upper limit to the aperture radius of 4.5$^{\prime\prime}$
was imposed for the few occasions when the image FWHM's were greater
than 2.0$^{\prime\prime}$, to avoid any contamination from a faint
visual companion 6.5$^{\prime\prime}$ SSE of J0644+3344.

There was occasional cirrus during several of the nights when
J0644+3344 was observed.  The effect on the relative photometry was
minimal, however, due to our rather long exposures and the selection
of fairly bright reference stars.  Differential light curves for one
of the comparison stars relative to the mean of all the others
indicates that the typical 1\,$\sigma$ noise in the J0644+3344 light
curves was 0.001 to 0.003 magnitudes whenever the extinction due to
clouds was less than 0.1--0.2 mags.  For cloud extinctions up to 1.0
magnitudes, the noise in the relative light curves increased to
approximately 0.01 mag.  The observations were halted whenever there
was a drop of more than one magnitude due to clouds.  

Stars in the field of J0644+3344 were calibrated using Landolt (1992)
UBVRI standards during the clear nights of 06~Dec 2005 and 05~Mar
2006.  A description of the photometric solutions can be found in
Appendix~I.  Table~3 contains a list of the comparison stars used and
their magnitudes.  The absolute photometry of the comparison stars and
the derived extinction coefficients for Mt.\ Bigelow were used to
correct all of the J0644+3344 light curves to the standard system.

\begin{table} 
\caption{Log of Kuiper 1.55m Photometric observations.}
\label{Table 2}
\centering
\begin{tabular}{lccccc}
\hline\hline
UT date & HJD Start & HJD End & Filters & N  & Inst. \\
\hline
02 Feb 05    &3403.63100   &3403.86903   &  R         &   300         &    4K\\
04 Feb 05    &3405.61762   &3405.87962   &  R         &   319         &    2K\\
01 Mar 05    &3430.60707   &3430.80255   &  R+B       &   203         &    2K\\
02 Mar 05    &3431.59528   &3431.83146   &  R+B       &   256         &    2K\\
03 Mar 05    &3432.60205   &3432.79780   &  R+B       &   209         &    2K\\
04 Nov 05    &3678.81874   &3679.03925   &  R         &   338         &    4K\\
05 Nov 05    &3679.78777   &3680.03462   &  B         &   343         &    4K\\
06 Nov 05    &3680.85072   &3681.03434   &  B         &   255         &    4K\\
05 Dec 05    &3709.71712   &3710.04482   &  R+B       &   429         &    4K\\
06 Dec 05    &3710.71717   &3711.01314   &  R+B       &   305         &    4K\\
07 Dec 05    &3711.70199   &3712.01251   &  R+B       &   406         &    4K\\
08 Dec 05    &3712.69835   &3713.03623   &  R+B       &   445         &    4K\\
06 Jan 06    &3741.63524   &3741.29503   &  R+B       &   412         &    4K\\
07 Jan 06    &3742.62904   &3742.95755   &  U+I       &   410         &    4K\\
10 Jan 06    &3745.60197   &3745.95607   &  V+R       &   452         &    4K\\
13 Oct 06    &4021.91698   &4022.02386   &  B         &   109         &    4K\\

\hline
\end{tabular}
\end{table}

\begin{table*} 
\caption{J0644+3344 Field Reference Stars}
\label{Table 3} 
\centering
\begin{tabular}{llllllllll}
\hline\hline
Object  &    RA  & DEC  &         V   &    U-B  &   B-V &    V-R  &   V-I  &  n&  m\\
  &  (J2000)  & (J2000)  &        &     &   &     &    &  obs. &  nights\\
\hline
J0644    &06:44:34.4 &+33:44:57    &13.564$\pm$0.243   &-0.732$\pm$0.065  & 0.279$\pm$0.053  & 0.274$\pm$0.057  & 0.780$\pm$0.099  &   3    &   2\\
									       	     
J0644-A  &06:44:20.4 &+33:40:06    &11.599$\pm$0.006    &0.375$\pm$0.006  & 0.823$\pm$0.001  & 0.397$\pm$0.034  &   -      	   &   2    &   1\\
									       	     
J0644-B  &06:44:57.0 &+33:40:29    &14.219$\pm$0.002    &0.141$\pm$0.008  & 0.766$\pm$0.001  & 0.436$\pm$0.001  &   -              &   2    &   1\\
									       	     
J0644-C  &06:44:33.0 &+33:41:12    &11.925$\pm$0.006    &0.817$\pm$0.003  & 1.113$\pm$0.006  & 0.506$\pm$0.040  &   -   	   &   2    &   1\\
									       	     
J0644-D  &06:44:34.1 &+33:42:07    &14.704$\pm$0.023    &0.108$\pm$0.001  & 0.656$\pm$0.006  & 0.360$\pm$0.008  & 0.705$\pm$0.013  &   3    &   2\\
									       	     
J0644-E  &06:44:25.9 &+33:42:53    &13.594$\pm$0.022    &0.497$\pm$0.005  & 0.934$\pm$0.005  & 0.491$\pm$0.006  & 0.972$\pm$0.011  &   3    &   2\\
									       	     
J0644-F  &06:44:27.1 &+33:44:23    &15.923$\pm$0.018    &0.039$\pm$0.006  & 0.436$\pm$0.007  & 0.253$\pm$0.011  & 0.542$\pm$0.020  &   3    &   2\\
									       	     
J0644-G  &06:44:15.3 &+33:44:57    &14.043$\pm$0.019    &0.107$\pm$0.016  & 0.661$\pm$0.014  & 0.372$\pm$0.014  & 0.708$\pm$0.020  &   1    &   1\\
									       	     
J0644-H  &06:44:37.1 &+33:45:17    &15.529$\pm$0.006    &0.097$\pm$0.010  & 0.660$\pm$0.013  & 0.369$\pm$0.008  & 0.750$\pm$0.015  &   3    &   2\\
									       	     
J0644-I  &06:44:29.1 &+33:45:19    &14.499$\pm$0.014    &0.959$\pm$0.009  & 1.170$\pm$0.009  & 0.590$\pm$0.010  & 1.159$\pm$0.017  &   3    &   2\\
									       	     
J0644-J  &06:44:30.0 &+33:45:51    &15.619$\pm$0.013    &0.160$\pm$0.001  & 0.744$\pm$0.010  & 0.409$\pm$0.014  & 0.816$\pm$0.024  &   3    &   2\\
									       	     
J0644-K  &06:44:20.2 &+33:45:58    &15.567$\pm$0.010   &-0.058$\pm$0.012  & 0.559$\pm$0.013  & 0.309$\pm$0.005  & 0.656$\pm$0.009  &   3    &   2\\
									       	     
J0644-L  &06:44:59.1 &+33:46:13    &13.298$\pm$0.002   &-0.007$\pm$0.008  & 0.647$\pm$0.004  & 0.347$\pm$0.002  &   -              &   2    &   1\\
									       	     
J0644-M  &06:44:33.9 &+33:46:20    &14.767$\pm$0.004   & 0.227$\pm$0.003  & 0.765$\pm$0.008  & 0.418$\pm$0.008  & 0.826$\pm$0.014  &   3    &   2\\
									       	     
J0644-N  &06:44:27.6 &+33:46:55    &14.006$\pm$0.004   & 1.619$\pm$0.005  & 1.465$\pm$0.010  & 0.752$\pm$0.008  & 1.449$\pm$0.014  &   3    &   2\\
									       	     
J0644-P  &06:44:57.2 &+33:47:46    &13.679$\pm$0.001   & 1.095$\pm$0.012  & 1.220$\pm$0.002  & 0.598$\pm$0.006  &   -          &   2    &   1\\
									       	     
J0644-Q  &06:44:17.7 &+33:47:43    &14.162$\pm$0.007   & 0.066$\pm$0.008  & 0.658$\pm$0.014  & 0.334$\pm$0.013  & 0.703$\pm$0.022  &   3    &   2\\
									       	     
J0644-R  &06:44:34.8 &+33:47:52    &13.026$\pm$0.004   &-0.020$\pm$0.007  & 0.562$\pm$0.003  & 0.298$\pm$0.006  & 0.604$\pm$0.011  &   3    &   2\\
									       	     
J0644-S  &06:44:47.4 &+33:48:33    &12.972$\pm$0.003   & 0.000$\pm$0.013  & 0.381$\pm$0.001  & 0.196$\pm$0.008  & 0.416$\pm$0.013  &   3    &   2\\
\hline
\end{tabular}
\end{table*}

\subsection{Spectroscopic and Spectropolarimetric Observations with the 2.3m Bok Telescope} 
Spectroscopic observations of J0644 covering multiple orbital
cycles were obtained on the  Steward Observatory 2.3m Bok telescope,
located  on Kitt Peak, during 2005 January 16-17, 2005 December
04-07, and 2006 January 09 (see Table 3).  The optical spectra were obtained
using the Boller and Chivens Spectrograph at the Ritchey-Chretien f/9
focus using multiple grating settings along with a 1200$\times$800 15 $\mu m$ pixel CCD.
For 2005 January observations, we used a 1st order 600 line/mm grating blazed at 3568 \AA.  
With a 2.5 arcsec slit width, a typical spectral
resolution of 5.5 \AA\ was achieved at a reciprocal dispersion of 1.87
\AA/pixel on the CCD.  Typical exposure times of 300 seconds yielded
characteristic S/N ratios of $60-70$.   The 2005 December and 2006
January observations were taken with the 832 line/mm grating and used
in 2nd order at two different grating tilts to cover the blue region
($\sim$ $3800-4850$ \AA) and 1st order in conjunction with a UV filter
to cover the red region ($\sim$ $5000-7000$ \AA) with two separate
grating tilts.  A 1.5 arcsec slit was used producing a typical
spectral resolution of 1.8 \AA.  Before each observation, the
instrument was rotated to align the slit perpendicular to the horizon,
minimizing the effects of atmospheric dispersion.

Standard IRAF routines were
used to reduce the data.  The wavelength calibration was established
with He-Ar arc-lamp spectra, interpolated between exposures taken
before and after each observation,  to account for any small
wavelength shifts that may occur while the telescope tracks an object.
The spectra were flux calibrated using the Massey et al.\ (1988)
spectrophotometric standards G191-B2B and BD+28$^{\circ}$ 4211,
observed over a range of zenith angles covering the program stars'
airmass range.

Spectropolametric observations of J0644 were also taken with the 2.3m
Bok telescope using the instrument SPOL (Schmidt et al.\ 1992) on
2005 Dec 31.  The observations covered the region from $4200-8400$
\AA\  with a resolution of 16 \AA.  A 2 hour interval between binary phases
$0.45-0.75$ (as determined below) were observed.  These observations, however, showed no
significant signs of circular polarization with a polarimetric
fraction of +0.003$\pm$0.003\%.  The lack of polarization would appear to rule out
J0644 as being a polar or intermediate-polar type of CV as the typical magnetic fields
would have easily been detected.

\begin{table} 
\caption{Log of Spectroscopic and Spectropolarimetric Observations with the 2.3m Bok Telescope}
\label{Table 4}
\centering
\begin{tabular}{lcccc}
\hline\hline
Date  & MHJD Start  &  MHJD End     & Spec. Range  & N \\
\hline
2005 Jan. 16        &          &3387.2321          &   Blue     &  1    \\
2005 Jan. 16        &3388.2599 &3388.3950          &   Blue     &  4    \\
2005 Dec. 04        &3709.2061 &3709.5283          &   Blue     & 38    \\
2005 Dec. 05        &3710.2056 &3710.5262          &   Red      & 37    \\
2005 Dec. 06        &3711.1840 &3011.5262          &   Blue     & 46    \\
2005 Dec. 07        &3712.1995 &3012.5306          &   Red      & 40    \\
2005 Dec. 31        &3735.2401 &3735.3214          &   Spol     & 9     \\
2006 Jan. 09        &3745.1532 &3745.4544          &   Red      & 28    \\                           
\hline
\end{tabular}
Blue: 3880 - 5040 \AA; Red: 5500 - 6200 \AA; Spol: spectropolametric
\end{table}

\begin{figure}
  \includegraphics[width=6.5cm,angle=90]{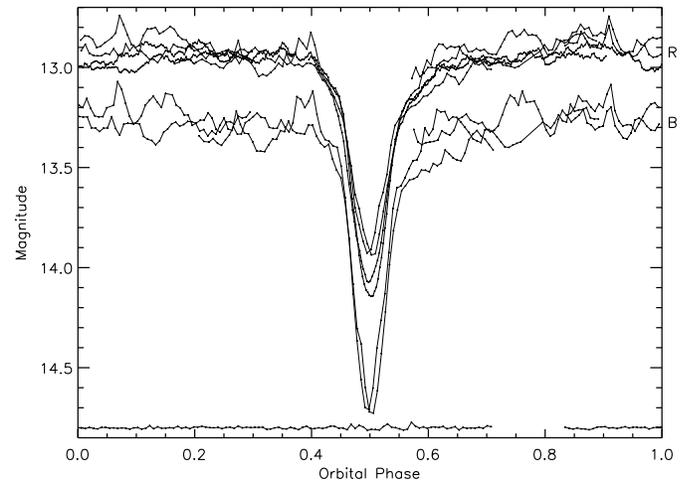}
  \caption[J0644.Photm.ps]{R and B photometry showing the eclipse of the primary star.  The points at the bottom 
                                 of the plot are the R residuals for the 2005 March 02 data which characterize the standard
                                 differential photometry errors.}
\end{figure}

\section{Analysis}
\subsection{Photometry}
Photometric coverage in the B and R bands show deep 1 to 1.2 magnitude
primary eclipses of the hot primary star (see Fig.\ 2).  The eclipse
lasts around 1.3 hours and has a characteristic ``V'' shape.  There
also is considerable 'flickering' in the light curve, indicative of
mass transfer, but no evidence of a secondary eclipse.
With multiple night coverage  of J0644, the fundamental orbital
period of the system was subsequently determined to be $P$= 6.4649808$\pm$
0.0000060 hrs.  We define phase 0.0 to be the red-to-blue crossing of
the primary star, such that primary eclipse happens at phase 0.0.  The
current ephemeris for primary eclipse is given by:
\begin{equation} T_{\mathrm{mid-eclipse}} = HJD(2453403.62501) + (0.26937431) \times E,  \end{equation}
with associated errors of $\pm$0.00022 days in T$_0$ and $\pm 2\times10^{-7}$ days in $P$. 

Although there is a strong He II $\lambda$4686 emission line, part of which is usually
associated with a ``hot spot'' on the accretion disk, we see very little, if any, photometric evidence for a ``hot spot'' modulation in the
light curve.
The lack of a noticeable ``hot spot'' modulation in the light curve, directly before eclipse,
is typical of NL CVs.  The large, bright accretion disk, combined with the fact that the stream material
does not fall deep into the white dwarf potential well before it hits the disk edge, combine to make
a low contrast of the ``hot spot'' with respect to the approximately constant out of eclipse light from the accretion disk.

During the course of our observations (Tables 1-4) as well as the time period covered by the ROTSE observations
(2 October 1999 to 29 March 2000) no dwarf nova-type outburst has been observed.  The constant out of eclipse 
brightness and lack of outburst
observations is consistent with J0644 being a NL type CV.

\subsection{Spectroscopy}
Spectroscopically, the out-of-eclipse observations show a strong blue
continuum, on which are superimposed broad emission lines due to H I and He I
which have a deep, narrow absorption component not easily seen at
lower resolution, Fig.\ 3.  At higher resolution, however, the
absorption component is resolved, see Fig 4.  The strong emission
line, He II $\lambda$4686, has no resolved absorption feature and requires a
very high level of excitation at its formation site in the accretion disk.  Note that strong
He II lines occur in both polars (highly magnetic CVs) and in high $\dot{M}$, longer period
CVs. About 1/3 of the CVs known with strong He II $\lambda$4686
emission lines are non-magnetic (Szkody et al.\ 1990).

\begin{figure}
  \includegraphics[width=6.6cm,angle=90]{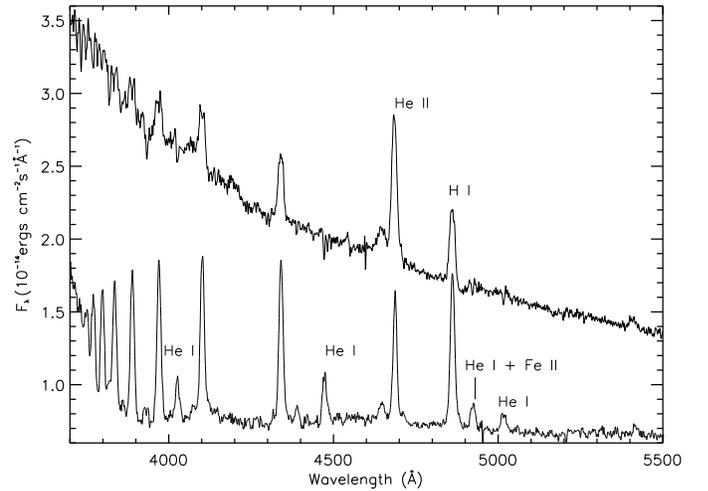}
  \caption[DavidSing.fig2.ps]{Optical spectra of J0644+3344 both in eclipse (bottom) and out (top), showing
    the H I Balmer series, He I, and He II as well as possible Fe II (Mason \& Howell 2005). A 'Bowen fluorescence' feature is also visible at $\lambda$4640 \AA\, caused by ionized
    carbon and nitrogen.}
\end{figure}

\begin{figure}
  \includegraphics[width=6.6cm,angle=90]{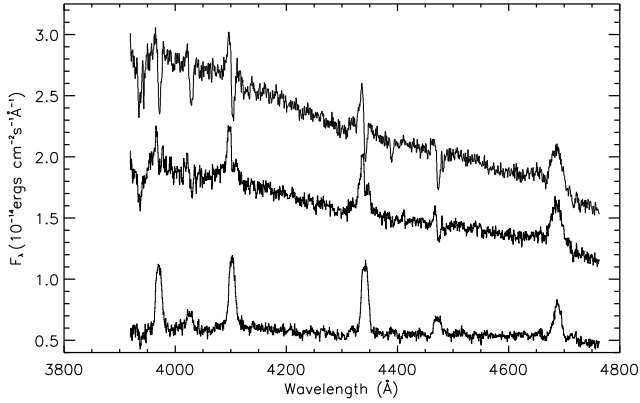}
  \caption[eclipse.ps]{Optical spectra of J0644+3344 before primary eclipse (top), during ingress (middle), and at
                          total primary eclipse.  Notice, compared to Fig.\ 3 the absorption lines are fully resolved and
                          can easily be seen.}
\end{figure}

\subsection{Line Flux}
The He II $\lambda$4686 emission line flux is seen  to modulate on
time scales equal to the orbital period, with its value dramatically
decreasing during primary eclipse.  This would argue that most of
the He II emission originates from an inner disk line forming region near the
primary star and not in an outer accretion disk hot-spot (see Fig.\
5).  The emission is not entirely eclipsed, however, suggesting that
at least part of the He II emission arises outside of a small inner disk line forming
region and/or in the hot-spot.  Furthermore, the FWHM of the He II line is seen to  decrease
by $\sim$70\% during eclipse as would be expected when emission from
an inner accretion disk with higher Keplerian velocity is eclipsed,
decreasing the emission line width.

\begin{figure}
\includegraphics[width=6.6cm,angle=90]{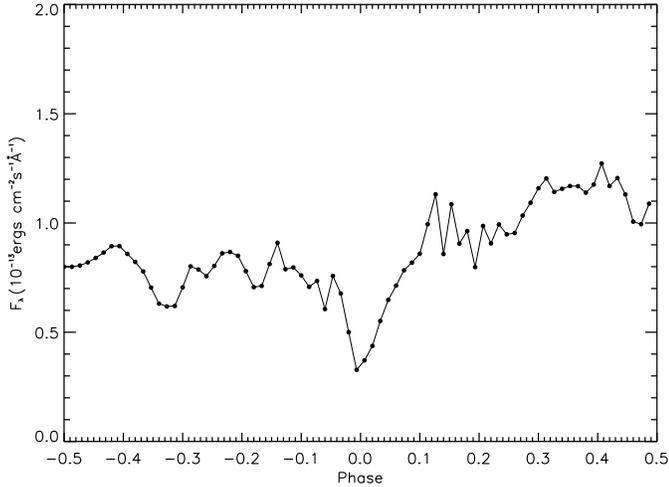}
\caption[flux_plot_ave_HeII.ps]{2005 Dec 04 \& 06 average line flux from the He II $\lambda$4686 emission line.  The flux is seen 
                                 to be at a minimum during primary eclipse at phase 0.0.  This would suggest most
                                 the emission comes from a hot inner line forming region.}
\end{figure}

A flux curve of the H I emission line flux for H$\gamma$ is seen in Fig.\ 6 
produced by averaging the Dec 04 and 06 2005 data.
This line, while out of eclipse, is comprised of a broad emission line
component associated with the accretion disk and a narrow absorption
component which can reach well below the continuum (see Fig.\ 4 \&
Fig.\ 7).   The line flux is summed over both the absorption and
emission line components, making it possible to track the full emission component's
relative contribution to the flux during primary eclipse.  The flux also
provides information on the component's line formation location.  When
the eclipse begins at phase $-$0.15, point A in Fig.\ 6,   the
H$\gamma$ flux is seen to abruptly decrease, implying the emission
line forms in the outer accretion disk.   The flux continues to
decrease up until phase $-$0.07, point B, at which point the flux is
observed to dramatically  increase up to phase $-$0.02, point C.
During this ingress period of the eclipse, between points B and C, the
central accretion region \& primary star are being eclipsed, thus  hiding
the absorption line component and increasing the H$\gamma$ flux.  The  spectra
change from a blue continuum dominated by deep absorption cores, to a
fainter flat continuum containing only emission lines at phase
$-$0.02, point C.   At this phase, the H I emission is seen to be at
its maximum and the primary star and inner accretion disk are
eclipsed.  The narrow  absorption core returns at phase
0.03, point D, lowering the flux which decreases until phase 0.09,
point E,  when the primary eclipse ends.  The time between points B \&
C is 20.7 minutes, which we use to estimate the radius of the
absorption line region (see \S3.6).

\begin{figure}
\includegraphics[width=6.6cm,angle=90]{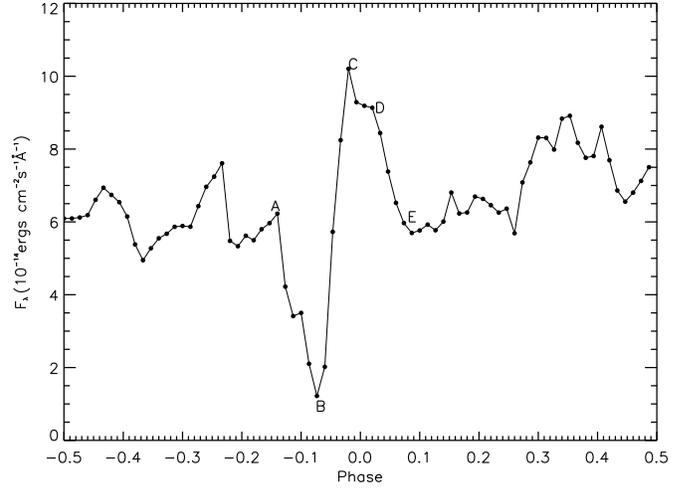}
\caption[flux_plot_ave_HI.ps]{Average line flux from the H$\gamma$ line using the 2005 Dec 04 \& 06 data.}
\end{figure}

The Balmer decrement provides information about the temperature and
density of the emission line forming gas as well as its opacity.  
It is defined as the ratio between any two H I line intensities,
D$_{\nu}(H\alpha/H_{\beta})=I_{\nu}(H\alpha)/I_{\nu}(H\beta)$.  When
determining the Balmer decrement from spectra, where fluxes are
measured, the direct comparison of flux-ratios to  intensity-ratios
assumes that the measured emission lines have the same profiles (see
Mason et al.\ 2000 and references therein).   The Balmer emission
lines of J0644 are complex due to the absorption line
components, changing the line profile in  a non-uniform manner.  The
emission lines in eclipse, however, have no absorption line cores and
have similarly broadened profiles, making a measurement of D$_{\nu}$
more reliable.  Measuring the flux-ratio centered on primary
eclipse gives  a value of D$_{\nu}(H\gamma/H\beta)=0.91$.  Williams
(1991) modeled the Balmer decrement for H I emission lines from
accretion  disks using grids of temperature, density and inclination.
Although D$_{\nu}$ alone does not provide a unique solution for the
accretion disk temperature or density, it can provide valuable ranges
for those parameters.  Seen in Fig.\ 8, our eclipse measured Balmer decrement
corresponds to a number density range between log N$_{o}$=$12.8-13.8$
and a temperature range between 8,000 and 15,000 K.  The presence of
He II in the accretion disk would suggest that an upper temperature range
between 10,000 and 15,000 K is more realistic for some portions of the disk, 
indicative of a hot, NL accretion disk.  Using this upper
temperature range, the line forming region density is limited to log N$_{o}$=$12.8-12.9$.

\begin{figure*}
  \centering
\includegraphics[width=10cm,height=17.5cm,angle=90]{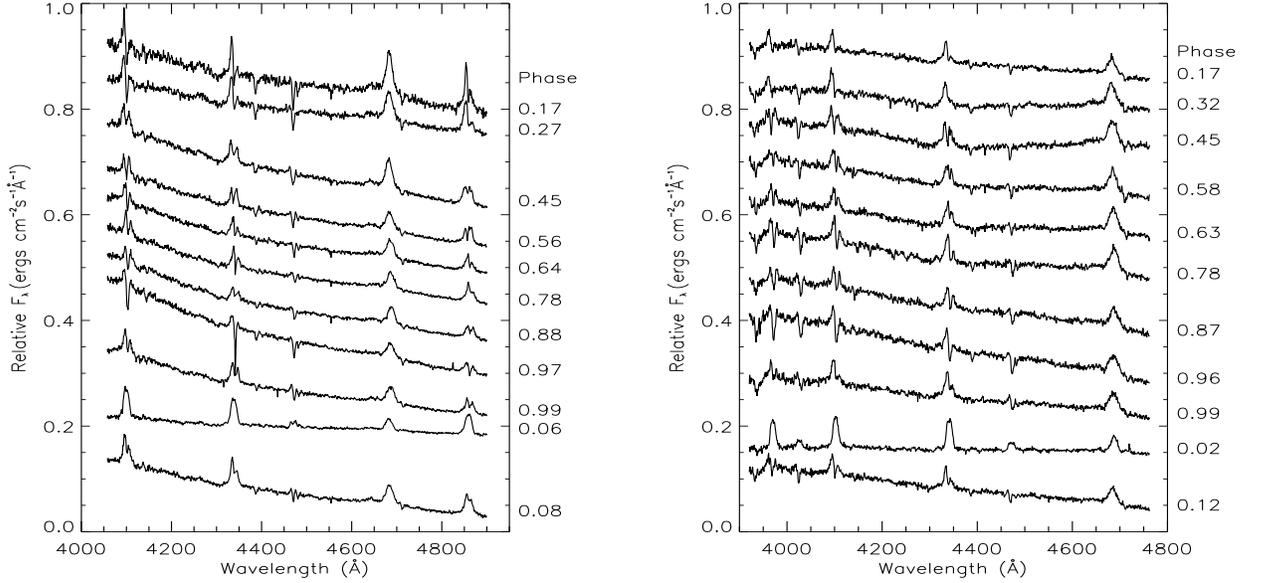}
  \caption[spectra_sequence.ps]{2005 December 04 (left) and 06 (right) spectra of J0644+3344.}
\end{figure*}

\begin{figure}
\includegraphics[width=6.6cm,angle=90]{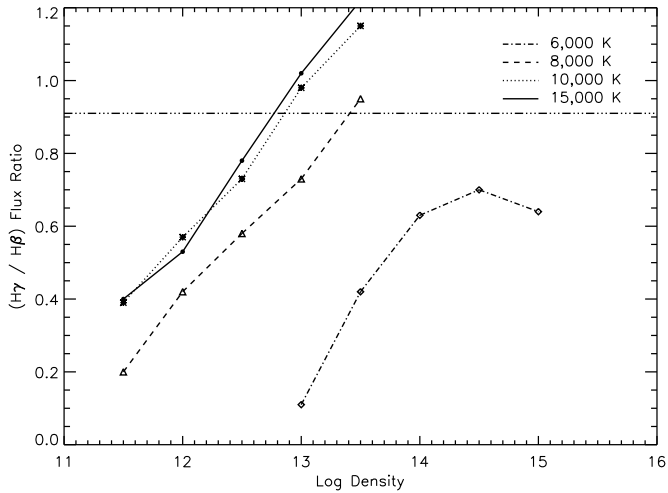}
\caption[BalmerDecFig.ps]{Balmer decrement (H$_{\gamma}$/H$_{\beta}$) vs. log density (g cm$^{-3}$) for four
                          different temperatures calculated by Williams (1991).  The values are for
                          an inclination of 80$^{\circ}$, except for 6,000 K which is calculated at 52$^{\circ}$.  The
                          value measured for J0644+3344 (H$_{\gamma}$/H$_{\beta}$=0.91) is shown as a horizontal line.}
\end{figure}

\subsection{Primary Star Radial Velocities}
Radial velocity measurements were performed on the H I and He I
absorption lines as well as the H I, He I, and He II emission lines
using the 2005 December and 2006 January 2.3m Bok telescope observations.  The
spectra were velocity shifted to a heliocentric rest frame where the
HJD of the midpoint of each exposure was calculated.    The radial
velocities were measured using a center-of-wavelength,
$\lambda^{\prime}$, for each line profile  to calculate a Doppler
velocity.  For each line profile of interest, $\lambda^{\prime}$ was
estimated following the formula, $\lambda^{\prime}=\frac{\sum \lambda
f_{\lambda}}{\sum f_{\lambda}}$.   We assumed a circular orbit and then fit the radial velocity
curves with a sine function of the form,
\begin{equation}  V(t) =  \gamma  + K \sin{\left[ 2\pi \frac{(t-T_{0})}{P}-2\pi\phi \right]},   \end{equation}
using a gradient-expansion algorithm to compute a non-linear least squares fit.  The fitted parameters
were; the system velocity $\gamma$ , the stellar Keplerian velocity $K$, and
the phase offset$\phi$.  The photometric ephemeris was used, where $\phi_0=0.0$ corresponds to primary eclipse.

The absorption lines of H I, He I, and Mg II (see Fig. 9) are seen to phase with
the primary star.   The sine fits over the entire period produce $K$ velocities between
$150-180$ km s$^{-1}$ for the absorption lines.  Between phases $-$0.5
to 0, the velocity curve appears sinusoidal, producing better sinusoid fits and larger amplitudes, but then remains
approximately constant between phases 0.0 and 0.5 (see Fig 10 and Table~5).  There
appears to be no clear reason for the flat velocity curve, but
curiously, the secondary star's velocity curve exhibits the same
phenomena (see \S 3.5).  A few possible causes of a flat velocity curves may be emission and absorption line blending, magnetic effects, non-uniform disk emission, or effects from an accretion stream.

The emission lines of H I, He I, and He II are also seen to phase with
the primary star.  The radial velocity curve derived from the He II
emission line is seen in Fig 11.  This line is single-peaked and is 
sinusoidal throughout the orbital period, possibly providing a much more
precise measure of the $K$ velocity than the H I and He I emission 
lines.  The He II velocity curve gives a $K_{\mathrm{He
II}}$=151$\pm$5 km s$^{-1}$ and $\gamma$=-96$\pm$3 km s$^{-1}$.  
The He II emission line appears, therefore, 
to be the only velocity feature that is observed to behave
sinusoidally throughout the entire orbital phase.  As we have seen, the bulk of the He
II emission arises in the inner-accretion disk which should
accurately follow the center of mass of the primary star.  No
significant phase lags are observed in any of the emission or
absorption line velocity solutions (see Table 5).  We use the He II velocity
solution, $K_{\mathrm{He II}}$, as the best fit for the $K$ velocity of
the primary star, $K_{P}$.

\begin{table} 
\caption{J0644+3344 radial velocities fitting over the entire phase (top half) and over phases -0.5 to 0 (bottom half).}
\label{Table 5}
\centering
\begin{tabular}{lllll}
\hline\hline
Feature         & $K$          & Phase Offset   & $\gamma$    & $\chi_{\nu}^2$\\
 & km s$^{-1}$ & $\phi-\phi_{phot}$ & km s$^{-1}$ &\\
\hline
He~I $\lambda$4026    &   189$\pm$6  &    0.038$\pm$0.005  &     26$\pm$4   &     4.25\\
He~I $\lambda$4471    &   145$\pm$2  &    0.043$\pm$0.002  &      5$\pm$1   &     12\\
Mg~II $\lambda$4481   &   169$\pm$4  &    0.045$\pm$0.004  &     32$\pm$3   &     4.93\\
He~II $\lambda$4686   &   150$\pm$4  &    0.043$\pm$0.004  &    -95$\pm$3   &     1.22\\
\hline
He~I $\lambda$4026    &   209$\pm$20 &    0.033$\pm$0.006  &    11$\pm$16   &   2.27\\
He~I $\lambda$4471    &   266$\pm$8  &    0.026$\pm$0.002  &  -73$\pm$6     &   5.84\\
Mg~II $\lambda$4481   &   257$\pm$17 &    0.015$\pm$0.004  &  -29$\pm$13    &   4.18\\
He~II $\lambda$4686   &   167$\pm$14 &    0.039$\pm$0.007  &  -104$\pm$10   &  1.06\\
\hline
\end{tabular}
\end{table}

The best fit $\gamma$ velocities, measured from both the emission and absorption lines,
give widely different values ranging from $\sim$-100 to 25 km s$^{-1}$.
These differences are similar to those typically seen in CVs such as the dwarf novae WZ Sge 
(Mason et al.\ 2000) and VY Aqr (Augusteijn 1994).  The velocity differences would suggest that the
accretion disk emission is non-symmetric, which leads to the inconsistencies
observed.

\begin{figure}
  \includegraphics[width=6.6cm,angle=90]{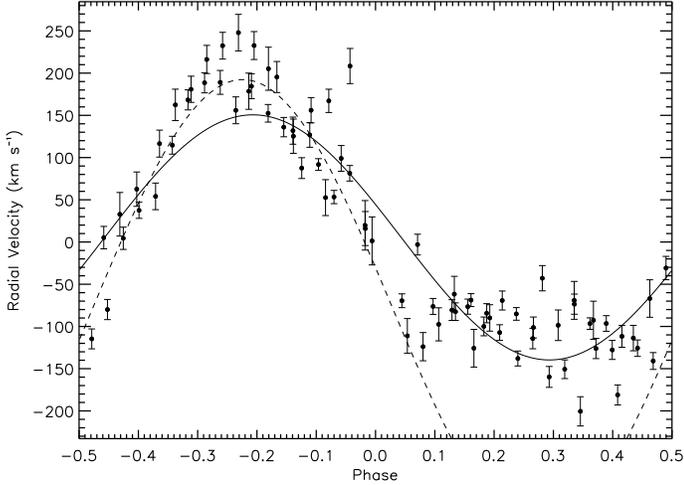}
  \caption[J0644.RV.HeI.4478.ps]{Radial velocity curve of the He I $\lambda$4471 \AA\ absorption line.  Overplotted is
a sinusoidal fit over the entire period (solid line), and using only phases -0.5 to 0 (dashed line) which produces a better fit.}
\end{figure}

\begin{figure}
  \includegraphics[width=6.6cm,angle=90]{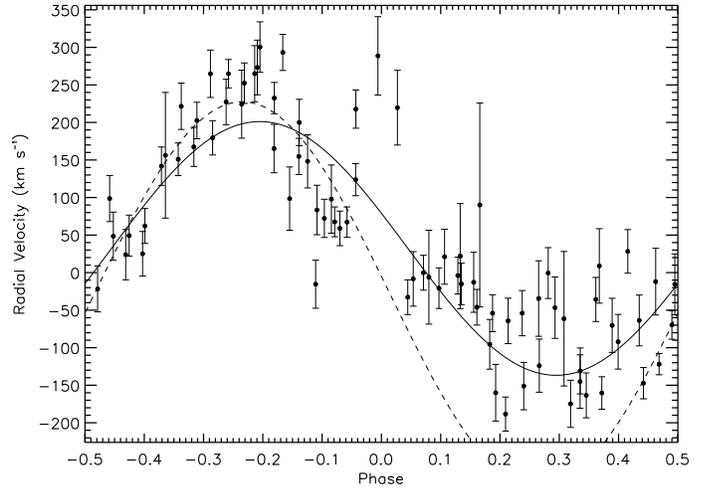}
  \caption[J0644.RV.MgII.ps]{Radial velocity measurements of J0644+3344 using the Mg II absorption line plotted against the 
orbital phase.  The velocity variations of Mg II phase with the primary star. Overplotted is
a sinusoidal fit over the entire period (solid line), and using only phases -0.5 to 0 (dashed line) which produces a better fit.}
\end{figure}

\begin{figure}
  \includegraphics[width=6.6cm,angle=90]{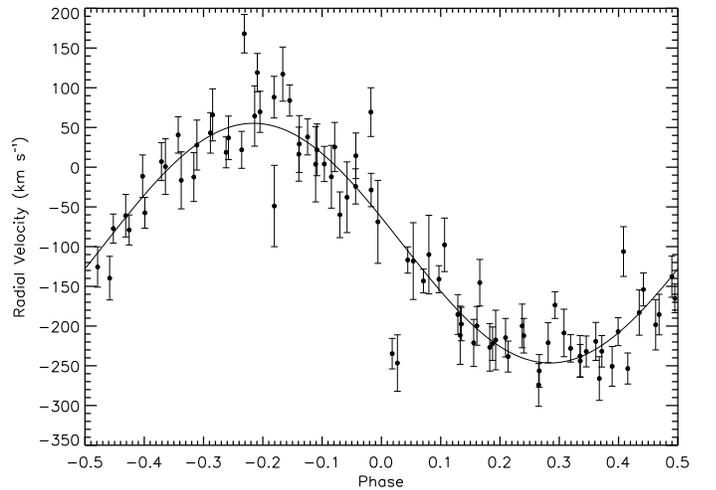}
  \caption[J0644.RV.HeII.ps]{Radial velocity measurements of J0644+3344 using the He II emission line.}
\end{figure}

\subsection{Secondary Star Radial Velocities}
The 2005 December 07 and 2006 January 09 observations, taken with a
first order 1200/mm grating blazed at 5346\AA, were optimized to
detect the weak spectral signature of the late type secondary and
measure its radial velocities using a cross-correlation method.
J0644's 2MASS H-K color suggested that the secondary might be a K
star.  K stars have numerous strong absorption lines in
our chosen wavelength range, 5035--6195\AA, in particular the Mg
triplet near 5100\AA, whereas J0644 exhibited an almost pure
continuum spectrum, except for one He emission line at 5876\AA.  
Given that the secondary star features are week, we note that the observed
spectral region used here is well to the blue of where the CCD starts to show fringing near 6800\AA, so our
cross-correlations can't be affected by incompletely removed fringes,
as might occur in the red.

Along with the J0644+3344 observations, we observed a number of main
sequence star templates at very high S/N, including several radial
velocity standards and MK spectrum standards ranging from F7V to
M0V.  For the January run, the grating was slightly tilted in order to
shift the spectra by 5\AA\ ($\sim$5 pix) relative to the December
observations, and special care was taken to observe template stars
with the widest possible spread in radial velocity.  These precautions
ensured that the combined super-template for the cross-correlation
would be nearly free of any residual calibration features, after the
individual templates were shifted to the rest velocity and median
filtered.  The spectral standards enabled us to independently verify,
and in some cases rederive, the spectral types of the other main
sequence stars, using standard spectral typing techniques.

The cross-correlation procedure was performed as follows.  First, the
spectra from both nights were dispersion-corrected onto the same
logarithmic wavelength scale.  We fit the continuum for each spectrum,
divided by the fit, and then subtracted 1.0 from the result, in order
to get a continuum value of zero.  All of the cross-correlations used
the double-precision version of the IRAF FXCOR task, with a ramp
filter function, and a gaussian fit to the cross-correlation peak.
The filter cuton and cutoff parameters were optimized for highest
sensitivity to narrow lines at the observed 2.3\AA\ resolution.  The
high S/N main sequence spectra were cross-correlated against the
radial velocity standards, then all were shifted to the rest velocity.
The error of the mean for the velocity zero point is 4~km~s$^{-1}$.
Next, the J0644 spectra were cross-correlated against each of the
available template spectra in the wavelength range 5040-5800\AA\ +
5900-6185\AA\ (avoiding the He 5875 emission line and the interstellar
Na D).  The strongest correlations throughout the orbital cycle were
always found to be the fits to the K3--K5 spectral types.  The final
super-template was created by combining 14 spectra with types K0--K7.
When the J0644 spectra were cross-correlated one last time
against the super-template, most of the correlations were quite
strong (Tonry-Davis factors of 5 to 12), confirming the choice of our
observed wavelength range for this K-type secondary.

The resulting radial velocities and 1$\sigma$ error
bars\footnote{FXCOR calculates velocity errors that are relatively
correct, but include an unknown multiplicative factor.  We tried
several Monte Carlo simulations to estimate the size of the true
internal errors, and found that they were always small compared to the
scatter of the points around the fitted curve.  Slit-centering errors
are most likely responsible for a large part of the external error.
When compared to the final sinusoidal fit, however, the fxcor errors
were observed to closely match an expected 1$\sigma$ error
distribution and were thus adopted as our velocity errors.} can be
seen in Fig.~12.  The solid curve is a sinusoidal fit to the
datapoints between phases $-$0.2 and +0.4 (using the known photometric
orbital period), where all the velocity correlations were strong and
the secondary can be clearly seen in the spectra.  The velocity correlations become
quite weak, and in a couple of cases are indistinguishable from the
noise, between phases 0.4 and 0.5, when the secondary is approaching
the point where it is eclipsed by the primary.  Interestingly, the
correlations from phase $-$0.5 to $-$0.25 are about as strong as those at
phase 0.0, but they do not follow
the expected orbital motion of the secondary.  Rather, it appears as
though the measured velocities are at least partly due a hot, dense stream of
material approaching us more rapidly than the secondary, although the
best correlation fits are still with early K spectra.

A fit using only those data points that exhibit sinusoid variation,
and that are likely to come from the secondary star, results in a
$K$ velocity of $K_{2}$=192.8$\pm$5.6 km s$^{-1}$ and a system gamma
velocity of $\gamma$=-7.1$\pm$1.3 km s$^{-1}$.  As stated above, no
obvious trends with changes in the secondary's spectral type with
orbital phase were observed.

\subsection{Determination of the Binary System Component Parameters}
The binary star parameters can be estimated using the eclipse as a
constraint on the orbital inclination.  With the orbital period and $K$
velocities determined in \S3.4 \& 3.5, the mass of the primary, $M_{P}$,
becomes a function of the inclination, $i$, given by,
\begin{equation}  M_{P}(i) = \frac{PK_{2}(K_{P}+K_{2})^{2}}{2\pi G \sin^{3}i }, \end{equation}
where G is the gravitational constant and the other parameters are
listed in Table 1.  The minimum mass of the primary corresponds to
$i=90^{\circ}$ giving $M_{P}(90^{\circ})=0.632 M_{\odot}$.  The mass
ratio, measured from the radial velocity curves and independent of
inclination assuming circular orbits, is found to be $q$=1.285.   A minimum $M_{P}$ along with
$q$, give a minimum secondary mass of $M_{2}(90^{\circ})=0.492
M_{\odot}$.   The orbital separation, $a$, then comes from  Kepler's
third law given by,
\begin{equation}  a^3(i) = G[M_{P}+M_{2}]\left[\frac{P}{2\pi}\right]^{2},    \end{equation}
resulting in a minimum separation of $a_{min}$=1.82 $R_{\odot}$.

The inclination can be constrained by the eclipse geometry with
estimates for the effective size of the primary 
(star+disk), the cooler secondary star radius, and the semi-major axis of the binary.  A minimum inclination angle
corresponds to one which still produces primary eclipses.  By
effective size of the primary, we mean the aggregate size of the
primary star and the inner accretion  disk which dominates the blue
optical continuum and is observed to eclipse.  The primary's effective
size is determined in \S4 to be $\sim$0.31 $R_{\odot}$ while the
secondary radius can be estimated by the volume radius of its Roche
lobe to be 0.78 $R_{\odot}$, using $q$ and $a_{min}$.
These radii estimates constrain the inclination to $i>76^{\circ}$.  The resulting masses are therefore
determined to fall in the range
M$_{P}=0.632-0.692$ M$_{\odot}$ for the primary
and M$_{2}=0.492-0.538$ M$_{\odot}$ for the secondary.  The uncertainty
in the absolute masses is dominated by the inclination uncertainty.
Although detailed eclipse modeling may ultimately provide a better
estimation of the inclination, the masses of {\it each} component are
determined in this study to $\sim$4\%, an accuracy level rarely
achieved in CVs.

\begin{figure}
  \includegraphics[width=6.6cm,angle=90]{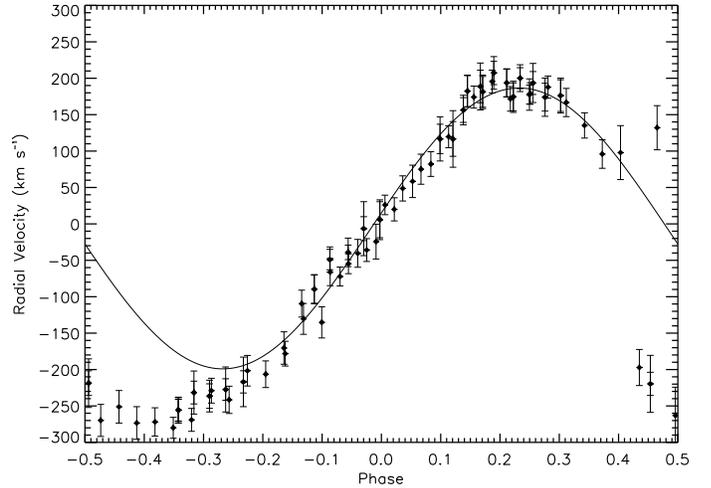}
  \caption[J0644.RV.k4.all.ps]{Radial velocity of the secondary star derived from cross-correlation of the continuum
between H$\alpha$ and H$\beta$.  A sine-curve fitted between the phases of -0.2 and 0.4, where the velocity variation appears sinusoidal, is overplotted.}
\end{figure}


\section{Discussion}

In cataclysmic variables, the primary star spectrum can sometimes be obtained
by subtracting a spectrum taken just before primary eclipse with one
taken at mid-eclipse when the primary is fully eclipsed.  This technique has
previously been used to extract a white dwarf spectrum for eclipsing CVs such as
Z Cha (Marsh et al.\ 1987).   On our 2005 December 04 run, 
two spectra were obtained such that the exposures
corresponded to phases just before and after the completion of
primary eclipse (see Fig.\ 4).  The first spectrum
was taken at phase $-$0.0788 while the second at phase $-$0.0177,
corresponding closely with points B and C in Fig.\ 6.    The
subtracted spectrum (see Fig.\ 13) shows a blue $\sim$25,000 K black body continuum
along with narrow absorption features from H I, He
I, Mg II $\lambda$4481, and Ca II H\&K.

The time it takes to fully eclipse the primary region can be used to
measure its diameter, assuming we know the orbital separation and
period.  The separation is measured in \S3.6 to be $\sim$1.85
R$_{\odot}$ and the orbital period is 0.26937420 days.  With these
parameters, the secondary star revolves around the primary, it its rest
frame, with a velocity of 348 km s$^{-1}$.  Measured from the emission
line profiles in \S3.3, the time it takes to eclipse the primary
region (the primary star and inner accretion disk) 
is 20.7 minutes, giving a radius of $\sim$0.31 R$_{\odot}$.

The narrow H I and He I absorption features in Fig 13 have a characteristic
Gaussian FWHM broadening of around $560-570$ km s$^{-1}$.  These
absorption line profiles clearly do not resemble a WD, being much to
narrow.  They are also narrower than the expected widths of subdwarf
or main sequence B stars.   Given the strong He II $\lambda$4686
emission, which out of eclipse is stronger than H$\beta$, it seems
likely that the Balmer line wings are partially filled in with He II
emission at $\lambda$3968, 4100, 4338, and 4859 which may contribute
to the narrowness of the lines.   No observed superhumps or phase lags
along with the eclipse characteristics of the absorption region makes
us  conclude that it must be confined to the inner accretion disk and
not a hot-spot, which is the case in other NL binaries such as
RW Tri (Groot et al.\ 2004).  This subtracted spectrum contains the
region with both the primary star and the  inner-accretion disk,
although it is currently unclear as to the relative contribution of
each component.  Given the narrowness of the lines, the spectra seems
to indicate that J0644 contains either an optically thick
inner-accretion disk, or that the primary star is not a WD but rather
some newly formed pre-WD.

\begin{figure}
  \includegraphics[width=6.6cm,angle=90]{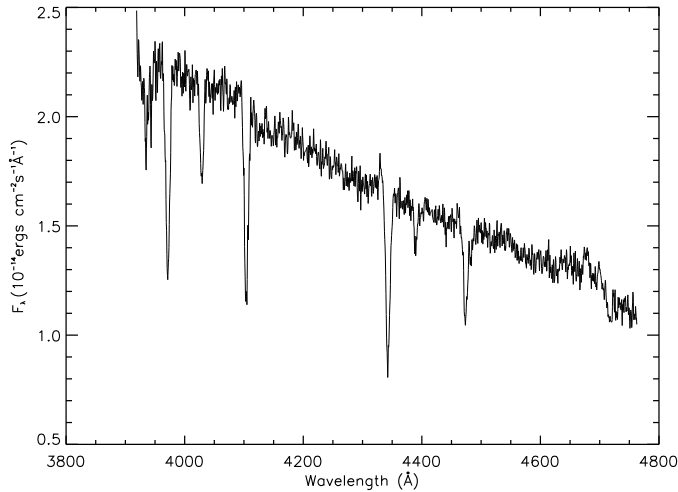}
  \caption[primary.ps]{Extracted spectrum of the region containing the inner accretion disk 
                          and primary star.  The spectrum is consistent with a $\sim$25000 K black body
                          which has strong narrow absorption lines from H I, He I, Mg II, and Ca II.}
\end{figure}

A white dwarf primary is likely, but requires a large hot optically
thick accretion disk which can hide both the  optical and X-ray flux
emanating from the WD.  Optically thick accretion  disks with single
black body temperatures in similar NL CVs have thus far been
observed to be much cooler than the 25,000 K temperature measured
here.  With an optically thick disk inward of 0.31 $R_{\odot}$, it is
unclear as to the origin of the  high velocity components of the
emission lines.  These components require  Keplerian velocities of
$\sim$1100 km s$^{-1}$ and should originate closer to the primary star
at  distances of around 0.1 $R_{\odot}$.   The dilemma is highlighted
by narrow absorption cores within wide emission lines.  In other
optically thick disks,  wide absorption lines are seen with narrow
emission line cores.  An accretion disk with both an
optically thick and thin region, however, could potentially explain
the observed continuum eclipse light curve.  The light curve shows a
broad shallow ``U''-shaped  eclipse profile toward the beginning and end of
the eclipse, and a much deeper ``V''-shaped curve at the center of the eclipse.  An outer
optically thin region could produce the ``U''-shaped eclipses as seen
in normal NL CVs, while an optically thick flared inner region
could produce the ``V''-shaped curve as seen during mid-eclipse
(Knigge et al.\ 2000).

An alternative explanation is to assume that the primary star is not
hidden and the star's photosphere can be seen with deep H I and He I
absorption lines and a 25,000 K black body temperature.  These
characteristics (along with the eclipse  light curve) suggests the
star is not a WD, as the absorption lines are much to narrow.   If not
a WD, the primary would most likely be a pre-WD or subdwarf (sdO or sdB), given its
characteristic size of 0.31 R$_{\odot}$  and mass 0.65  M$_{\odot}$.
These scenarios are complicated with the lack of any secondary
eclipses thus far observed, the extreme narrowness of the
absorption lines, and the high luminosity of subdwarf stars.  With 
the stated characteristics mass and size, the
primary star would have a surface gravity around  log g = 5.26, which
should produce absorption lines which have widths as wide as those
seen in subdwarf stars.  The lines, however, are significantly
narrower than typical subdwarf absorption lines, likely ruling out a
subdwarf for the primary star.  A lack of secondary eclipses (most
notably in the R band) would also point toward a compact primary as
very little of the secondary  star appears to be hidden during
eclipse.  Unfortunately the relative spectral energy distribution of
each  stellar component is unknown, thus it is difficult to place
limits on the size of the primary star and accretion disk, due solely
to a lack of secondary eclipses.  Observations will be needed in the
infrared,  where the secondary should be comparable or brighter to the
accretion disk and primary.  These observations could reveal secondary
eclipses and make it possible to estimate the  primary region's size
and shape.

There are several possibilities to help explain the extracted primary
spectrum and other characteristics of J0644 by attempting to fit
the binary into a known CV classification.  There are two obvious
NL CV  classifications which J0644 could correspond to.
The first being a {\it thick-disk} or UX UMa NL CV.
Although UX UMa type CVs exhibit a wide range of spectral features
(see Warner 1995), a main feature of these binaries are wide
absorption lines (typically in H I, He I, and Ca II) with narrow
emission line cores.  These CVs have been shown to contain optically
thick inner accretion disks, giving rise to the wide Doppler broadened
absorption lines.     An optically thin outer disk then provides the
narrow emission component.  This spectral characteristic, however, is
the opposite of what is observed in J0644 which has a narrow
absorption component in the center of wide emission lines.
Furthermore, for J0644 the inner accretion disk temperatures
would have to be $\sim$ 25,000 K.  At this high temperature, accretion
disks are expected to be optically thin, produce X-rays, and would not
produce the absorption lines observed.  

The second possibility is that J0644 belongs to the SW Sex
NL class of CVs.  SW Sex stars are eclipsing NL binaries
which show peculiar features interpreted as stream-disk overflows,
disk winds, and flared accretion disks.  One of the spectral features
of these stars are narrow central absorption components in H~I and He~I,
which look a lot like the features seen in J0644.  However, this
central absorption component in SW Sex stars only appears around phase
0.5 and can appear and disappear quickly.  These characteristics have
been explained by an overflowing stream resulting in absorption which
can only be viewed through a flared disk during phases $0.2-0.6$.  In
J0644, however, the absorption components  persist throughout the
orbital cycle, only disappearing during primary eclipse.  This would
seem to rule out any overflows in J0644 as the absorption is
constrained to the inner disk/primary star region and is nearly always
visible.  SW Sex stars also show single peaked emission lines whereas
other eclipsing CV binaries show double peaked lines expected from  a
Keplerian disk.  This single peaked nature has been suggested as at
least partly due to a flared disk which produces a flat-topped
emission line profile.  Although J0644 does not show single
peaked emission from H I and He I (except during eclipse),  the He II
emission line is observed to be single peaked, possibly indicating
that it too has a flared disk.  SW Sex stars also show large radial
velocity phase lags relative to the photometric ephemeris and
wind-formed P Cygni profiles, traits not observed in J0644.  The
emission lines of J0644 are significantly narrower than SW Sex.
Dhillon et al.\ 1997 measured a FWZI (full width at zero intensity) of
5700 km s$^{-1}$ for He II in SW Sex which compares to $\sim$2200 km
s$^{-1}$ for J0644.


\section{Conclusions}

J0644 is a bright new example of an eclipsing binary with ongoing
mass transfer.   Spectroscopic and photometric observations have
revealed J0644 to be a young cataclysmic variable binary
dominated by either an optically thick inner accretion disk or a white
dwarf progenitor.   J0644 is a unique binary that does not easily
fit into any known cataclysmic variable subtypes, although it most
closely resembles a UX Uma or SW Sex type NL CV.
With no observed magnetic fields, P Cygni profiles, phase lags, or superhumps 
there is no evidence with which to evoke the usual physical interpretations.  These include 
truncated accretion disks, winds, and stream-disk hot spots.  Numerical models and further
observations in different wavelengths regimes will ultimately be needed to fully understand this unique system
and its properties.  Given its bright nature and multitude of 
features observed in this binary, J0644 will become an important binary for many future studies.

\begin{acknowledgements}
We wish to thank For, BiQing (University of Texas) in her observing help during four 61'' nights 
and the referee Ed Guinan for his helpfull comments.  D.K.S. is supported by CNES.
This publication makes use of the data from the
Northern Sky Variability Survey created jointly by the Los Alamos
National Laboratory and University of Michigan. The NSVS was funded by
the US Department of Energy, the National Aeronautics and Space
Administration and the National Science Foundation.  This publication
makes use of data products from the Two Micron All Sky Survey, which
is a joint project of the University of Massachusetts and the Infrared
Processing and Analysis Center/California Institute of Technology,
funded by the National Aeronautics and Space Administration and the
National Science Foundation.

\end{acknowledgements}

\begin{appendix}

\section{Photometric Solutions}

We observed the field around J0644 on two clear nights, 
UT 06 Dec 2005 and 05 Mar2006.  UBVR observations were obtained in
December for 18 Landolt standard stars (when the I filter was not
available), and UBVRI observations of 14 Landolt standards were
obtained in March.  The photometric solutions for the two nights were:

\begin{equation}  \mathrm{U} = u - 1.0490 + 0.1115*(u-b) + 0.0284*Xb - 0.4918*Xu    (0.0434) \end{equation}

\begin{equation}  \mathrm{B} = b + 0.4055 + 0.0096*(b-v) + 0.0015*Xv - 0.2557*Xb    (0.0135) \end{equation}

\begin{equation}  \mathrm{V} = v + 0.3480 - 0.0246*(b-v) + 0.0059*Xb - 0.1587*Xv    (0.0124) \end{equation}

\begin{equation}  \mathrm{R} = r + 0.2013 - 0.0423*(v-r) + 0.0066*Xv - 0.1279*Xr    (0.0152) \end{equation}
on 06 Dec 2005 and

\begin{equation}  \mathrm{U} = u - 1.3932 + 0.0939*(u-b) + 0.0246*Xb - 0.4354*Xu    (0.0441) \end{equation}

\begin{equation}  \mathrm{B} = b + 0.1057 - 0.0012*(b-v) - 0.0002*Xv - 0.2619*Xb    (0.0118) \end{equation}

\begin{equation}  \mathrm{V} = v + 0.0726 - 0.0427*(b-v) + 0.0112*Xb - 0.1672*Xv    (0.0122) \end{equation}

\begin{equation}  \mathrm{R} = r - 0.1210 - 0.0826*(v-r) + 0.0137*Xv - 0.1153*Xr    (0.0106) \end{equation}

\begin{equation}  \mathrm{I} = i - 1.2931 + 0.0251*(v-i) - 0.0042*Xv - 0.0735*Xi    (0.0179) \end{equation}
on 05 Mar 2006, where the numbers in parentheses are the standard deviations of the fits.

The coordinates and the mean magnitude and colors for J0644 and
18 reference stars in the surrounding field are given in Table 3.  
The listed errors are the standard error of the mean (except
for J0644-G, for which we had only one measurement), although, in many
cases, the the systematic transformation errors are comparable or even
larger than the standard errors.

\end{appendix}


\begin{thebibliography}{}
\bibitem[Augusteijn94]{Augusteijn94}Augusteijn T. 1994, A\&A, 292, 481
\bibitem[Clarke (2002)]{Clarke}Clarke, D. 2002, A\&A, 386, 763
\bibitem[Dhillon]{Dhillon}Dhillon, V.~S., Marsh, T.~R.,\& Jones, D.~H.~P. 1997, MNRAS, 291, 694
\bibitem[Groot et al. (2004)]{Groot04}Groot, P. J., Rutten, R. G. M., \& Paradijs J. van 2004, A\&A, 417, 283
\bibitem[Hoard et al. (2002)]{Hoard}Hoard, D. W., Wachter, S., Clark, L. L., Bowers, T., P. 2002, ApJ, 565, 511
\bibitem[Knigge et al. (2000)]{Kin00}Knigge, C., Long, K. S., Hoard, D. W., Szkody, P. \& Dhillon V. S. 2000, \apj, 539, L49
\bibitem[landolt]{landolt}Landolt, A.~U. 1992, AJ, 104,340
\bibitem[Marsh et al. (1987)]{Marsh87}Marsh, T. R., Horne, K. \& Shipman, H. L. 1987, MNRAS, 225, 551 
\bibitem[Mason et al. (2000)]{Mason00}Mason, E., Skidmore, W., Howell, S. B., Ciardi, D. R., Littlefair, S. \& Dhillon, V. S. 2000, MNRAS, 318, 440
\bibitem[Mason et al. (2005)]{Mason05}Mason, E. \& Howell, S. B. 2005, A\&A, 439,301
\bibitem[Massey et al.(1988)]{Massey88}Massey, P., Strobel, K., Barnes, J. V., Anderson, E. 1988, \apj, 328, 315
\bibitem[Schwarzenberg-Czerny (1989)]{Schwarzenberg}Schwarzenberg-Czerny, A. 1989, MNRAS, 241, 153
\bibitem[Sing (2005)]{Sing05}Sing, D. K. 2005, PhD Thesis
\bibitem[Szkody et al. (1990)]{Szkody}Szkody, P., Garnavich, P., Howell, S., \& Kii, T. 1990, in 
Proceedings of the 11th North American Workshop on 
Cataclysmic Variables and Low Mass X-Ray Binaries, Edited by Christopher W. Mauche, Cambridge University Press, p.251
\bibitem[Schmidt et al. (1992)]{Schmidt92}Schmidt, G. D., Stockman, H. S., \& Smith P. S. 1992, \apj, 398, L57
\bibitem[Walker]{Walker}Walker, M. F. \& Herbig, G. H. 1954, \apj, 120, 278
\bibitem[Warner]{Warner}Warner, B. 1995 {\it Cataclysmic variable stars}, Cambridge Astrophysics Series, 
Cambridge, New York: Cambridge University Press
\bibitem[Williams (1991)]{Williams91}Williams, G. A. 1991, \aj, 101, 1929
\bibitem[Wozniak et al. (2004)]{Wozniak}Wozniak, P. R., Vestrand, W. T., Akerlof, et. al. 2004, AJ, 127, 3043
\bibitem[Zacharias et. al. (2004)]{Zacharias}Zacharias, N., Urban, S. E., Zacharias M. I., et al. 2004, AJ, 127, 3043

\end{thebibliography}
\end{document}